\documentclass[aps,prx,reprint,superscriptaddress,amsmath,amssymb,longbibliography]{revtex4-2}
\usepackage{bm,mathrsfs,booktabs,microtype,graphicx}
\usepackage[hidelinks]{hyperref}
\newcommand{\dd}{\mathrm{d}}
\newcommand{\Ccal}{\mathcal{C}}
\newcommand{\kf}{\mathrm{kf}}
\usepackage{comment}

\begin{document}

\title{Effective Contact Theory for Exotic Loosely Bound States in Strongly Interacting Expanding Matter}
\author{Xiaofeng Wang}
\email{xiaofeng_wang@ustc.edu.cn}
\affiliation{Department of Modern Physics, University of Science and Technology of China, Hefei, Anhui 230026, China}
\author{Zebo Tang}
\email{zbtang@ustc.edu.cn}
\affiliation{Department of Modern Physics, University of Science and Technology of China, Hefei, Anhui 230026, China}
\author{Zhangbu Xu}
\email{zxu22@kent.edu}
\affiliation{Physics Department, Kent State University, Kent, Ohio 44242, USA}
\affiliation{Physics Department, Brookhaven National Laboratory, Upton, New York 11973, USA}
\author{Chi Yang}
\email{chiy@cern.ch}
\affiliation{Key Laboratory of Particle Physics and Particle Irradiation (MOE), Institute of Frontier and Interdisciplinary Science, Shandong University, Qingdao, Shandong 266237, China}
\author{Wangmei Zha}
\email{wangmei.zha@cern.ch}
\affiliation{Department of Modern Physics, University of Science and Technology of China, Hefei, Anhui 230026, China}

\begin{abstract}
Short-distance correlations generate universal relations that can be
independent of microscopic details. Tan's contact is a prominent example,
connecting close pairs, high-momentum constituents, and bound-state formation
across atomic, condensed-matter, and nuclear systems. Whether an analogous
universality governs the nonequilibrium freeze-out of relativistic QCD matter is
an open question. We develop an effective contact theory that relates the
production of loosely bound states to continuum two-particle correlations in
heavy-ion collisions. The construction replaces the hard relative-momentum
cutoff of conventional coalescence by a regulated Bethe--Peierls kernel:
freeze-out localization fixes its momentum scale, while the composite yield
fixes its contact residue. Bound and continuum observables then become
projections of a common two-particle density matrix rather than independent
phenomena. Coulomb-bound $K\mu$ atoms provide an ideal realization
because their atomic, production, and source scales are widely separated. A contact-theory analysis of STAR $d\Lambda$ correlations supplies a
first bound--continuum consistency test and demonstrates how inferred
near-threshold parameters can depend on the finite expanding source. The
framework establishes a general strategy for relating coalescence,
correlations, and exotic bound states in strongly interacting expanding
matter.
\end{abstract}

\maketitle

\section{Introduction}

The production of loosely bound states in an expanding QCD medium poses a fundamental problem of scale.  Their constituents are emitted from a hot, compact source only a few femtometers across, whereas the final composite may extend over a much larger distance and have a binding energy far below the temperature of the medium.  Relativistic heavy-ion collisions provide a unique laboratory for this subject: they produce light nuclei, hypernuclei, candidate hadronic molecules, and potentially exotic
Coulomb-bound atoms despite conditions seemingly unfavorable to their survival.  This tension is often described as the ``snowball in hell'' puzzle
\cite{Sun:2022xjr,BraunMunzinger2015Snowballs,
Oliinychenko:2018ugs,Chen:2018tnh}.  Pion--muon atoms have been observed in particle decays \cite{Coombes:1976hi,Aronson:1982bz}, and their production in ultrarelativistic collisions was considered by Baym~\cite{Baym:1993ae} and Kapusta~\cite{Kapusta:1998fh}. Baym and Braun-Munzinger studied how localization affects Coulomb corrections to two-particle correlations \cite{Baym:1996wk}. Wang~\cite{Wang:2024njn,Wang:2026ftb} recently proposed that localization and Coulomb attraction at kinetic freeze-out greatly enhance the production of $\pi\mu$, $K\mu$, and $p\mu$ atoms.

Coalescence and femtoscopy models describe complementary aspects of the same source. Coalescence relates composite yields to the local phase-space density of constituents \cite{Butler:1963pp,Scheibl:1998tk}, whereas femtoscopic correlations probe relative emission coordinates and final-state interactions \cite{Koonin:1977fh,Lisa:2005dd}. A third description is supplied by Tan's universal relations: when the two-body wave function has a short-distance $1/r$ singularity, a contact $C$ controls both bound-state normalization and a $k^{-4}$ momentum tail \cite{Tan:2008ypg,Tan:2005xvg,Braaten:2008uh,Braaten:2010if}. Braaten, Ingles, and Pickett~\cite{Braaten:2024cke} recently used the contact density at kinetic freeze-out to describe the production of weakly bound hadronic particles. These approaches differ in what they retain of the microscopic pair dynamics.
A sharp coalescence model selects a phase-space volume, a Wigner-overlap
calculation folds the emission function with the composite wave function,
and contact theory characterizes the unresolved pair by a boundary condition followed by an adiabatic evolution.

The application of a universal short-range contact to a Coulomb-bound atom is neither obvious nor trivial. The physical Coulomb potential is long ranged, and its regular $1s$ wave function does not have the Bethe--Peierls $1/r$ singularity. Nevertheless, Wang's mechanism attributes the production enhancement to Baym's localization at freeze-out, rather than to a late classical capture process extending over the Bohr radius. This suggests an effective description in which the localized production amplitude is short ranged even though the asymptotic eigenstate is Coulombic. The distinction between production and propagation is central to our construction.

This use of an effective theory is not just  convenient for a model treatment.  Condensed-matter physics provides many precedents in which the microscopic interaction between charged constituents is Coulombic, while the measured low-energy excitations are quasiparticles governed by local effective interactions.  Screening, collective response, restricted kinematics, and coarse grain reorganize the microscopic dynamics into observable poles, residues, response functions, and universal coefficients \cite{Braaten:2010if}.  These effective quantities are physical within their domain of resolution even though they are not parameters of the bare Coulomb potential.  We apply the same logic to a localized freeze-out source: its effective pair pole and contact can appear in atom yields and two-particle correlations.

The proposed procedure is as follows: 
\begin{itemize}
\item     
Wang's construction supplies two empirical inputs: a localization momentum $p_0$ and an integrated atom yield. 
\item 
We replace the hard coalescence sphere by a smooth wave-function kernel that preserves those two inputs. 
\item 
Tan's boundary condition identifies the form of that kernel and converts its two matched parameters into an effective scattering length $a$ and contact $C$. Thereby, Tan's universality provides a universal completion of the otherwise discontinuous coalescence prescription. The same kernel predicts the short-distance relation of the constituent's relative momenta. 
\end{itemize}
Comparison with a normalized experimental correlation requires its absolute pair normalization and control of long-range Coulombic final-state interactions. In $K\mu$, the atomic, production, and source scales are separated. The atom
yield fixes the matched normalization, while the continuum provides an independent test of the momentum dependence.

\section{Baym's Localization}

For a kaon and muon with reduced mass
$ m_r=\frac{m_Km_\mu}{m_K+m_\mu}\simeq86.99~\mathrm{MeV}$, Wang's condition with Baym's localization~\cite{Wang:2024njn,Baym:1993ae} limits the relative kinetic energy to be less than the Coulomb potential at the characteristic freeze-out separation $r_0$, which provides a cutoff momentum scale: 
\begin{equation}
 p_0=\left(\frac{2m_r\alpha}{r_0}\right)^{1/2}.
 \label{eq:p0}
\end{equation}
In the framework of coalescence model, the atom spectrum is written as
\begin{equation}
 \frac{\dd N_{K\mu}^{A}}{\dd y\,\dd^2P_T}
 =B_2^{K\mu}f_K(x_KP_T)f_\mu(x_\mu P_T),
 \label{eq:wang_spectrum}
\end{equation}
where $x_i=m_i/(m_K+m_\mu)$. The total momentum $\bm P=\bm p_K+\bm p_\mu$ becomes the atom momentum in the bound channel and the constituent relative momentum of continuum: 
\begin{equation}
 \bm k\equiv\bm k^*
 =\frac{m_\mu\bm p_K-m_K\bm p_\mu}{m_K+m_\mu}. 
 \label{eq:relative_momentum}
\end{equation}
It follows: 
\begin{equation}
 B_2^{K\mu}=\frac{4\pi}{3}\frac{p_0^3}{m_r}.
 \label{eq:wang_B2}
\end{equation}
For $r_0=5$ fm, $p_0\simeq7.1$ MeV/$c$ and $B_2^{K\mu}\simeq1.7\times10^{-5}$ GeV$^2$. 

The source introduces a second momentum scale,
\begin{equation}
 \Lambda_{\rm fo}=\frac{1}{r_0}.
 \label{eq:freezeout_scale}
\end{equation}
For $r_0=5$ fm, $\Lambda_{\rm fo}\simeq39.5$ MeV/$c$. Wang's $p_0$ is therefore smaller than the inverse localization length, leaving a finite interval in which the pair may be insensitive to details below $r_0$. This scale separation, rather than the long-distance Coulomb spectrum by itself, motivates a contact description.

\subsection{The Coalescence Momentum Kernel}

Wang's model assigns a coalescence weight through a hard cutoff; other coalescence models use a Wigner-function overlap. The sharp prescription is economical for an integrated yield but does not predict a high-$k$ tail.

Our strategy is to replace, rather than supplement, the hard cutoff or the Wigner function integral by a smooth universal kernel,
\begin{equation}
 \Theta(p_0-k)\quad\Longrightarrow\quad
 \frac{C}{(k^2+a^{-2})^2}.
 \label{eq:replacement_strategy}
\end{equation}
The replacement preserves Wang's characteristic momentum and integrated
atom yield. These two matching conditions determine $a$ and $C$ in
Sec.~\ref{sec:matching}; the smooth kernel then supplies the relative-momentum
dependence.

\subsection{Localization as an Effective Short-range Interaction}
\label{sec: effective contact}
The physics foundation of the replacement is the localization of both constituent wave functions at kinetic freeze-out~\cite{Baym:1996wk}.  In Wang's mechanism~\cite{Wang:2024njn,Baym:1996wk}, the production-relevant kaon--muon overlap is concentrated at relative separations $r\lesssim r_0$.  A measurement with resolution longer than $r_0$ cannot distinguish the detailed Coulomb-enhanced overlap inside this region from a local two-body production with a thermalized system at freeze-out.  After the unresolved spatial structure is integrated out, the localized amplitude is therefore represented by an effective short-range interaction, even though the microscopic Coulomb potential remains long ranged. It produces the coordinate- and momentum-space windows
\begin{equation}
 r_0\ll r\ll a,
 \qquad
 p_0\ll k\ll\Lambda_{\rm fo}.
 \label{eq:dual_windows}
\end{equation}
Within these windows, both the source profile below $r_0$ and the asymptotic atomic wave function are not  resolved.  The long-wavelength amplitude can depend only on the effective length $a$ and on an overall close-pair strength.  This is precisely the setting in which a zero-range boundary condition~\cite{BethePeierls1935} and a contact~\cite{Braaten:2010if} provide the leading effective description.

The enhanced atom yield has a direct physics interpretation.  It indicates that localization produces a large close-pair amplitude with nonzero projection onto the bound channel.  At freeze-out resolution, that enhancement is equivalent to an effective short-range interaction supporting a correlated bound-channel state.  The word ``equivalent'' refers to equality of the low-resolution production matrix element, not to replacement of the microscopic Coulomb potential at all distances.

The effective state from the replaced kernel has a pole scale $p_0$ and energy
\begin{equation}
 k=i p_0,
 \qquad
 E_{\rm eff}=-\frac{p_0^2}{2m_r}
 =-\frac{\alpha}{r_0}.
 \label{eq:effective_pole}
\end{equation}
It must be distinguished from the physical Coulomb pole,
\begin{equation}
 k=i\gamma_C,
 \qquad
 E_C=-\frac{\gamma_C^2}{2m_r}.
 \label{eq:coulomb_pole}
\end{equation}
The effective pole characterizes localized production, whereas $\gamma_C$ characterizes the asymptotic atom.  Their separation permits a short-range production description and a long-range Coulomb final state to coexist without contradiction. When the reduced production amplitude contains a pole at $k=i p_0$, its position and residue control measurable quantities: $p_0$ fixes the turnover of the relative-momentum distribution and $C$ fixes both its integrated bound-channel strength and continuum amplitude. The effective pole is therefore a physical feature of the source-conditioned pair amplitude, although it need not be a pole of the vacuum two-body $S$ matrix.

\section{Universal kernel from Tan's Contact}
\label{sec:Universal Construction}
Tan's contact provides a universal description of particles that approach
one another within a distance much shorter than all other relevant length
scales. In this regime, the detailed interaction potential can be replaced
by a short-distance boundary condition, while its observable strength is
encoded in a single quantity, the contact $C$. Tan originally showed that
the same contact governs the number of close pairs, the $k^{-4}$
high-momentum tail, the interaction energy, and exact thermodynamic
relations \cite{Tan:2008ypg,Tan:2005xvg,Tan:2008huu}. These relations were subsequently
derived using quantum field theory and the operator-product expansion
\cite{Braaten:2008uh}, and their theoretical and experimental
implications have been reviewed extensively
\cite{Braaten:2004rn,Braaten:2010if}.
The underlying principle extends beyond ultracold atomic gases. Whenever
the many-body wave function factorizes at short separation into a universal
two-body function and a smooth function describing the remaining degrees of
freedom, the same short-distance coefficient can connect coordinate-space
pair correlations, momentum distributions, response functions, and
bound-state channels. Generalized contact formalisms have subsequently been
developed for nuclear systems, coupled channels, and finite-range
interactions \cite{Weiss:2015mba,Weiss:2016obx,
Weiss:2023kgy}. These developments establish contact as a
general effective-theory concept for strongly interacting systems.

For a large effective $S$-wave scattering length, Tan's boundary condition provides the wave function at small distance\cite{Tan:2008ypg,Tan:2005xvg,Braaten:2008uh}:
\begin{equation}
 \Psi(\bm R,\bm r)
 \underset{r\rightarrow0}{=}
 A(\bm R)\left(\frac1r-\frac1a\right)+O(r).
 \label{eq:tan_boundary}
\end{equation}
A shallow bound-state continuation is
\begin{equation}
 \Psi(\bm R,\bm r)=A(\bm R)\frac{e^{-r/a}}{r}.
 \label{eq:bound_completion}
\end{equation}
Its Fourier transform is
\begin{equation}
 \widetilde\Psi(\bm R,\bm k)
 =\frac{4\pi A(\bm R)}{k^2+a^{-2}}.
\end{equation}
Defining the integrated contact by
\begin{equation}
C=16\pi^2\int\dd^3R\,|A(\bm R)|^2,
 \label{eq:contact_definition}
\end{equation}
we obtain the direct universal kernel
\begin{equation}
\frac{\dd N_{\rm pair}}{\dd^3k}
 =\frac{1}{(2\pi)^3}
 \frac{C}{(k^2+a^{-2})^2}.
 \label{eq:universal_kernel}
\end{equation}

The full form is important near the turnover. For $k\ll a^{-1}$,
\begin{equation}
 \frac{C}{(k^2+a^{-2})^2}
 =Ca^4\left[1-2a^2k^2+O(k^4)\right],
 \label{eq:low_k_kernel}
\end{equation}
so the universal contribution is finite. It crosses smoothly to the contact tail rather than imposing a discontinuity at a coalescence boundary.
The kernel contains exactly the two quantities in Eq.~\eqref{eq:tan_boundary}: $a$ fixes the turnover and $C$ fixes the normalization. No auxiliary wave-function normalization is introduced. Its integral is
\begin{equation}
 \int\frac{\dd^3k}{(2\pi)^3}
 \frac{C}{(k^2+a^{-2})^2}=\frac{Ca}{8\pi}.
 \label{eq:kernel_integral}
\end{equation}

The high-$k$ expansion is
\begin{align}
 \frac{C}{(k^2+a^{-2})^2}
 &=\frac{C}{k^4}\left(1+\frac{1}{a^2k^2}\right)^{-2}\nonumber\\
 &=\frac{C}{k^4}-\frac{2C}{a^2k^6}
 +\frac{3C}{a^4k^8}+O(k^{-10}).
 \label{eq:high_k_expansion}
\end{align}
There is no $1/k^2$ term. The coordinate-space $1/r$ singularity produces a Fourier amplitude proportional to $1/k^2$; the probability, which is the squared amplitude, begins at $1/k^4$.

\section{Matching coalescence and Tan's contact parameters}
\label{sec:matching}

Matching the kernel turnover to Wang's scale in Eq.~\eqref{eq:p0} fixes the
effective length:
\begin{equation}
a=\sqrt{\frac{r_0}{2m_r\alpha}},
 \qquad \frac{1}{a}=p_0.
 \label{eq:a_from_r0}
\end{equation}
For $r_0=5$ fm, the effective freeze-out length is $a\simeq27.8~\mathrm{fm}$.
This matching gives the geometric relation
\begin{equation}
 \frac{1}{a}=p_0
 =\sqrt{2\gamma_C\Lambda_{\rm fo}},
 \qquad \gamma_C=m_r\alpha,
 \label{eq:scale_hierarchy}
\end{equation}
where $\gamma_C\simeq0.635$ MeV/$c$ is the physical Coulomb binding momentum. Numerically, $\gamma_C\ll p_0\ll\Lambda_{\rm fo}$. The effective scattering length is therefore intermediate between the freeze-out localization scale and the Coulomb Bohr radius.

Here $\gamma_C$ governs the asymptotic Coulomb state, $p_0$ the production
turnover, and $\Lambda_{\rm fo}$ the source resolution. Their separation
makes a differential continuum test possible beyond an integrated
coalescence yield~\cite{Braaten:2024cke}.

Matching Eq.~\eqref{eq:kernel_integral} to Wang's atom yield per unit
rapidity ($N_A\equiv\dd N_{K\mu}^{A}/\dd y$) gives
\begin{equation}
\begin{aligned}
 C&=\frac{8\pi}{a}N_A,\\
 &=8\pi p_0\frac{\dd N_{K\mu}^{A}}{\dd y}.
 \end{aligned}
 \label{eq:yield_contact_equivalence}
\end{equation}
This is the contact $8\pi/a$ of one universal shallow molecule multiplied by its multiplicity. 
Thus the enhanced yield is the integrated strength of the effective short-range bound-channel correlation.  A large yield implies a large contact in this leading-order representation.  This is stronger than an arbitrary refit because $p_0$ is independently fixed by localization and the functional form is fixed by the universal boundary condition.

To express the result through constituent spectra in Eq.~\eqref{eq:wang_spectrum}, and for a common nonrelativistic Boltzmann temperature, the normalized transverse spectra may be approximated by
\begin{equation}
 f_i(x_iP_T)=\frac{dN_i/dy}{2\pi m_iT_{\kf}}
 \exp\left(-\frac{p_T^2}{2m_iT_{\kf}}\right).
 \label{eq:boltzmann_spectrum}
\end{equation}
Substitution into Eq.~\eqref{eq:yield_contact_equivalence} and ~\eqref{eq:wang_spectrum} yields
\begin{equation}
 C=\frac{16\pi}{3m_r^2T_{\kf}}p_0^4
 \frac{\dd N_K}{\dd y}\frac{\dd N_\mu}{\dd y}.
 \label{eq:thermal_contact}
\end{equation}
For an expanding medium with blast-wave or non-equilibrium Tsallis-like spectra~\cite{Tang:2008ud,Chen:2020zuw}, the spectral overlap should be evaluated numerically. The contact scales as $C\propto p_0^4\propto r_0^{-2}$ at fixed spectra, whereas Wang's atom yield scales as $N^A\propto p_0^3\propto r_0^{-3/2}$.

The matching may also be performed differentially in atom transverse momentum ($P_T$):
\begin{equation}
\frac{\dd C}{\dd y\,\dd^2P_T}
 =\frac{8\pi}{a}
 \frac{\dd N_{K\mu}^{A}}{\dd y\,\dd^2P_T}.
 \label{eq:differential_contact}
\end{equation}
This form preserves the measured pair-momentum dependence while the kernel variable $k$ remains the constituent relative momentum.

\section{Distinction to the Braaten Construction}

Braaten, Ingles, and Pickett relate the final multiplicity of a weakly bound state to its contact density at kinetic freeze-out \cite{Braaten:2024cke},
\begin{equation}
 \frac{\dd N_A}{\dd y}
 =\frac{\kappa}{4\pi}
 \frac{\Ccal_{K\mu,\kf}}{n_{\pi,\kf}^{4/3}}
 \frac{\dd N_\pi}{\dd y}.
 \label{eq:braaten_formula}
\end{equation}
Combining Eqs.~\eqref{eq:yield_contact_equivalence} and \eqref{eq:braaten_formula} gives
\begin{equation}
 C=\frac{2\kappa}{a}
 \frac{\Ccal_{K\mu,\kf}}{n_{\pi,\kf}^{4/3}}
 \frac{\dd N_\pi}{\dd y}.
 \label{eq:C_braaten}
\end{equation}
The contact that normalizes the observed relative-momentum kernel is therefore the evolved image of the contact density at kinetic freeze-out.

At leading order in a virial expansion, the freeze-out contact density has the form~\cite{Braaten:2024cke}
\begin{equation}
 \Ccal_{K\mu,\kf}
 =\frac{16}{\pi}(m_rT_{\kf})^2z_Kz_\mu
 {\cal F}\left(\frac{a^{-1}}{\sqrt{2m_rT_{\kf}}}\right),
 \label{eq:virial_contact}
\end{equation}
where $z_K$ and $z_\mu$ are fugacities and ${\cal F}(w)\to1$ as $w\to0$. The dependence on $a^{-1}$ is weak in the universal limit. Therefore, a small binding momentum does not force the freeze-out contact to vanish; the abundance of close pairs is determined primarily by the thermal medium at the kinetic freeze-out. However, the phenomenological parameter $\kappa$ in Eq.~\eqref{eq:braaten_formula} and \eqref{eq:C_braaten} and scattering length enter the crossover condition and hence the conversion from freeze-out contact to
final yield. 

Coalescence, finite-source localization, and contact-based production
share a central feature: the freeze-out environment affects the
formation of loosely bound states. They differ in how they connect
that environment to the observed composite yield. Coalescence selects
constituents that are sufficiently close in phase space, either
through phenomenological coordinate- and momentum-space criteria or
through the overlap of their emission function with a bound-state
Wigner function \cite{Llope:1995zz,Scheibl:1998tk}. In the latter
formulation, the binding-dependent internal wave function and the
finite source enter the same projection. Describing the final
composite also requires a prescription for energy exchange with the
source or subsequent interactions.
Baym and Braun-Munzinger instead showed how Coulomb propagation maps
a spatially localized pair source onto measured continuum
correlations. Their semiclassical approximation applies under
specified conditions on the source and relative motion, while their
Coulomb-wave-function projection retains the finite source
explicitly \cite{Baym:1996wk}. Wang~\cite{Wang:2024njn} applies this
localization picture to Coulomb-assisted atom coalescence requiring formation of exotic atoms with a hard cutoff condition in Eq~\eqref{eq:p0} which depends on source radius and composite binding energy. Braaten, Ingles, and Pickett take a different route:
they estimate the contact density at kinetic freeze-out and propose
that it evolves with the expanding hadron gas until a crossover to a
dilute molecular population \cite{Braaten:2024cke}. Their crossover
criterion compares the pion spacing with the molecular size set by
the physical binding momentum, or equivalently the large scattering length. 
Our proposed approach extends Wang's construction by replacing the sharp
coalescence cutoff with a universal kernel at freeze-out, as summarized in
Eq.~\eqref{eq:replacement_strategy}. Like the framework of Braaten's construction, it assumes that an effective contact is established at
freeze-out. The two approaches differ, however, in how the contact is
determined. In the present construction, both the characteristic momentum
scale $p_0$ of the kernel and the contact strength are fixed by Wang's
localization mechanism, rather than by Braaten's virial expansion and the
subsequent evolution to a crossover density. These distinct assumptions
motivate a joint analysis of composite yields and constituent correlations,
which have long been recognized as complementary probes of the emission
source~\cite{Sun:2022xjr,Lisa:2005dd,STAR:2025jwe}.

\section{Prediction for Continuum Correlations}
\label{sec:continuum}

The smooth universal kernel replaces Wang's hard coalescence projector; it
does not replace the electromagnetic Hamiltonian. Localization makes the
production vertex unresolved on scales larger than $r_0$, whereas unbound
constituents can subsequently undergo long-range Coulomb propagation. These
operations act on the amplitude or density matrix and must not be added as
independent probabilities. The bound and continuum populations should
therefore be treated as projections of the same freeze-out density matrix.

The Coulombic unlike-sign attractive ($R_{K\pi}^{\rm US}(k)$) and like-sign repulsive ($R_{K\pi}^{\rm LS}(k)$) correlations are not perturbations around unity
in the first threshold bins. Experimentally, STAR measured charge-resolved
$K\pi$ correlations and observed the expected enhancement for the attractive
unlike-sign channels and suppression for the repulsive like-sign channels
~\cite{STAR:2003cqe}. Inspection of those published correlations shows that their
product is close to unity to 1\% level over the resolved low-momentum range. This empirical
cancellation motivates the
data-driven Coulomb-control observable
\begin{equation}
 {\cal P}_{K\pi}(k)=
 \frac{R_{K\pi}^{\rm US}(k)R_{K\pi}^{\rm LS}(k)}
 {B_{K\pi}^{\rm US}(k)B_{K\pi}^{\rm LS}(k)}
 \simeq1,
 \label{eq:kpi_charge_product}
\end{equation}
where the $B_s$ contain the sideband normalization, acceptance, purity, and
residual backgrounds. The product statement is our inference from the STAR
charge-resolved data, not an observable or fit result separately quoted by
STAR. Equation~\eqref{eq:kpi_charge_product} is therefore an empirical
finite-source and finite-bin relation, not an exact identity at mathematical $k=0$. Resolution-dominated threshold bins must
therefore be tested separately or excluded.

Under the minimal hypothesis that the next dominant correlation after Coulomb cancellation is the bound-continuum effect, the contact observable becomes
\begin{equation}
 {\cal E}_{\rm short}(k)=
 ({\cal P}_{K\mu}(k)-1)N_{K\mu}(k)
 =\frac{A_C}{(k^2+p_0^2)^2}.
 \label{eq:contact_observable}
\end{equation}
Here $N_{K\mu}(k)$ is the absolute correlated pairs in phase space per event without correlation and 
\begin{equation}
 A_C=\frac{C_{\rm fo}}{(2\pi)^3{\cal N}_{\rm pair}}
 \label{eq:correlation_contact_amplitude}
\end{equation}
converts the absolute pair-density contact to the normalization of the
experimental correlation. Atom production fixes $C_{\rm fo}$ and $p_0$, while
${\cal N}_{\rm pair}$ must be obtained from the absolute mixed-event pair
reference. A unit-normalized correlation alone cannot determine $A_C$. 
The raw unlike-sign enhancement is not by itself evidence for contact
universality. The proposed signal is a positive, yield-normalized remainder
in the charge product after the residual finite-source Coulomb product and any
allowed like-sign short-distance contribution have been constrained. Consequently, the
most stringent test is an absolutely normalized, joint yield--correlation
analysis: the yield fixes the bound projection, the mixed-event pair density
fixes the continuum normalization and tests
whether a single regulated source component accounts for both without an
additional matching coefficient.

\section{Extension to Searches for Loosely Bound States}

The same organization can be applied to searches for loosely bound hadrons in heavy-ion collisions.  Such searches commonly use either a composite yield or a two-particle femtoscopic correlation.  In an effective description these observables are complementary: the bound-state yield measures the integrated population of a pole channel, whereas the continuum correlation resolves the momentum dependence of the same low-energy pair amplitude~\cite{Lednicky:1981su,STAR:2025jwe}.

The central point is that effective-theory poles are not bookkeeping artifacts.  A pole of a vacuum scattering amplitude is observed through phase shifts, threshold line shapes, and bound-state energies.  A pole of a source-conditioned production amplitude is observed through its yield, turnover scale, and correlated-pair spectrum.  What differs is not whether the pole is real, but which Green function contains it and which experiment measures its residue.  Heavy-ion freeze-out supplies a finite-density, finite-size environment, so both vacuum poles and production poles may contribute to the same measured correlation.

For two constituents with reduced mass $\mu$, the low-energy $S$-wave amplitude may be written \cite{Bethe:1949yr}
\begin{equation}
 f(k)=\frac{1}{-a_{\rm phys}^{-1}+\tfrac12r_e k^2-ik}.
 \label{eq:effective_range_amplitude}
\end{equation}
A physical shallow bound state is a pole on the physical sheet at $k=i\gamma_B$, where
\begin{equation}
 -a_{\rm phys}^{-1}-\frac12r_e\gamma_B^2+\gamma_B=0,
 \qquad E_B=\frac{\gamma_B^2}{2\mu}.
 \label{eq:physical_bound_pole}
\end{equation}
In the zero-range limit, $\gamma_B\simeq a_{\rm phys}^{-1}$.  The pole position is a property of the two-body Hamiltonian or $S$ matrix; freeze-out normally controls how strongly this pole is populated, not whether it exists.

For a conventional shallow hadronic molecule, the leading contact contribution therefore has the form
\begin{equation}
 \Delta F_{12}(k)
 =\frac{1}{(2\pi)^3}
 \frac{C_{12,\rm fo}}{(k^2+\gamma_B^2)^2}.
 \label{eq:general_bound_kernel}
\end{equation}
If the same pole dominates the bound channel, its yield fixes the residue,
\begin{equation}
 \frac{\dd N_B}{\dd y}
 =\frac{C_{12,\rm fo}}{8\pi\gamma_B},
 \qquad
 C_{12,\rm fo}=8\pi\gamma_B\frac{\dd N_B}{\dd y}.
 \label{eq:general_yield_contact}
\end{equation}

The $K\mu$ construction presented in Sec.~\ref{sec: effective contact} represents a more unusual two-scale case.  The physical atomic pole is $\gamma_C$, whereas localization generates a broader effective production scale $p_0$.  Accordingly, a measured correlation may contain both a localized production term and a conventional final-state-interaction term. They coincide only when $p_0\simeq\gamma_B$.

Three regimes can therefore be distinguished:
\begin{align}
 \gamma_B\simeq p_0\ll\Lambda_{\rm fo}
 &:\quad \text{one physical-pole kernel},\nonumber\\
 \gamma_B\ll p_0\ll\Lambda_{\rm fo}
 &:\quad \text{two separated scales},\nonumber\\
 p_0\sim\Lambda_{\rm fo}
 &:\quad \text{no controlled zero-range window}.
 \label{eq:three_pole_regimes}
\end{align}
The first regime is the most direct realization of shallow-bound-state universality. The second is the regime proposed here for $K\mu$. The third requires explicit source profiles, effective-range terms, coupled channels, and other microscopic information.


\subsection{Source-Conditioned Scattering Parameters and an Apparent Pole}

A useful extension arises when the vacuum short-range interaction is attractive but does not support a bound state. Localization can then modify the repeated propagation of the pair sufficiently to produce a pole of the source-conditioned correlator. This possibility must be distinguished from an ordinary finite-source convolution, which changes the observed line shape but cannot by itself create a pole. Let $C_0$ denote the renormalized short-range coupling. In vacuum,
\begin{equation}
 T_{\rm vac}(E)=\frac{1}{C_0^{-1}-\Pi_{\rm vac}(E)},
 \label{eq:vacuum_tmatrix}
\end{equation}
and the absence of a physical-sheet zero of its denominator means that there is no vacuum bound state. In the localized freeze-out environment, the same coupling gives
\begin{equation}
 T_{\rm fo}(E;r_0)=
 \frac{1}{C_0^{-1}-\Pi_{\rm fo}(E;r_0)},
 \qquad
 \Pi_{\rm fo}=\Pi_{\rm vac}+\Delta\Pi_{\rm loc}.
 \label{eq:freezeout_tmatrix}
\end{equation}
No new microscopic force has been introduced: localization changes the two-particle propagator and hence the amount of repeated overlap sampled before decoupling. Expanding the localization correction at low momentum,
\begin{equation}
 \frac{2\pi}{\mu}\Delta\Pi_{\rm loc}(k;r_0)
 =-\lambda_0(r_0)-\lambda_2(r_0)k^2+O(k^4),
 \label{eq:localization_loop_expansion}
\end{equation}
defines the source-conditioned effective-range parameters
\begin{equation}
\begin{aligned}
 \frac{1}{a_{\rm fo}(r_0)}&=\frac{1}{a_{\rm vac}}-\lambda_0(r_0),\\
 r_{e,\rm fo}(r_0)&=r_{e,\rm vac}+2\lambda_2(r_0).
 \end{aligned}
 \label{eq:source_conditioned_ere}
\end{equation}
The signs in Eq.~\eqref{eq:source_conditioned_ere} follow the convention of Eq.~\eqref{eq:effective_range_amplitude}; the sign and magnitude of each $\lambda_i$ must be derived from, or fitted within, a specified source kernel.

An effective pole exists when
\begin{equation}
 C_0^{-1}-\Pi_{\rm fo}
 \left(-\frac{\gamma_{\rm fo}^2}{2\mu};r_0\right)=0.
 \label{eq:freezeout_pole_condition}
\end{equation}
In an effective-range representation this is equivalent to
\begin{equation}
 -a_{\rm fo}^{-1}(r_0)
 -\frac12r_{e,\rm fo}\gamma_{\rm fo}^2
 +\gamma_{\rm fo}=0.
 \label{eq:freezeout_effective_range_pole}
\end{equation}
The shallow solution and the binding energy that would be inferred from the freeze-out correlation are
\begin{equation}
\gamma_{\rm fo}=
 \frac{1-\sqrt{1-2r_{e,\rm fo}/a_{\rm fo}}}{r_{e,\rm fo}},
 \qquad E_{B,{\rm app}}(r_0)=\frac{\gamma_{\rm fo}^2}{2\mu}.
 \label{eq:apparent_binding_energy}
\end{equation}
As the source expands and dilutes, it may return to the continuum and need not emerge as a stable vacuum composite. 
Accordingly, parameters returned by a fit can mix the vacuum amplitude, a genuine freeze-out modification of the propagator, and source-model bias. A centrality analysis should compare pole, geometric, and vacuum hypotheses. Their characteristic source dependence is
\begin{equation}
\begin{aligned}
 \gamma_{\rm vac}&\propto r_0^0,
 &&\text{vacuum pole},\\
 p_0&\propto r_0^{-1/2},
 &&\text{Wang localization},\\
 k_{\rm geom}&\propto r_0^{-1},
 &&\text{source form factor}.
 \end{aligned}
 \label{eq:three_way_source_test}
\end{equation}
For a generic localized short-range interaction, the precise trajectory $\gamma_{\rm fo}(r_0)$ follows from Eq.~\eqref{eq:freezeout_pole_condition} and need not have the Wang exponent. The simultaneous centrality evolution of $\gamma_{\rm fo}$ and $C_{\rm fo}$ is therefore the decisive test of an emergent freeze-out pole.

The general research program is therefore to determine four quantities in the same collision classes: the composite yield, constituent correlation, source radius, and low-energy scattering parameters.  Agreement of the yield-normalized kernel with the continuum would show that coalescence and femtoscopy are observing the same pole contribution.  In this language, the bound-state yield measures its integrated strength, the correlation resolves its momentum dependence, and freeze-out determines the contact or residue with which the expanding source populates it.

\subsection{Existing Strange-Hadron Femtoscopy}

Existing searches illustrate both the reach and the ambiguity of this program.  The $\Lambda\Lambda$ correlation has been used to restrict the scattering-length--effective-range region associated with a possible $H$ dibaryon and to infer a model-dependent binding-energy interval \cite{ALICE:2019eol}.  STAR and ALICE $p\Omega^-$ measurements are sensitive to strong attraction and to bound-state scenarios, but their conclusions depend on source radii, spin-channel weights, Coulomb treatment, and the assumed interaction potential \cite{STAR:2018uho,ALICE:2020mfd}.  ALICE $p\Xi^-$ data establish attraction beyond Coulomb, while the existence and binding of a $p\Xi$ state require an additional analytic continuation and interaction model \cite{ALICE:2019hdt}.  The measured $\Lambda\Xi^-$ interaction is instead compatible with relatively small scattering parameters within current precision \cite{ALICE:2022uso}.

The $d\Lambda$ system is particularly instructive because the hypertriton is a shallow $d\Lambda$ bound state in its cluster description, while the neutral $d\Lambda$ pair is free of a Coulomb final-state interaction. STAR fitted correlation functions in three centrality classes simultaneously with a Lednick\'y--Lyuboshitz kernel: each class had its own Gaussian source radius, while the doublet and quartet scattering lengths and effective ranges were common. The spin weights were fixed to $1/3$ and $2/3$ \cite{STAR:2025jwe}. After correcting purity, weak-decay contamination, and hyperon feed-down, the fitted doublet parameters were analytically continued with
\begin{equation}
 -\frac{1}{f_0}=\gamma-\frac12d_0\gamma^2,
 \qquad B_\Lambda=\frac{\gamma^2}{2\mu_{d\Lambda}}.
 \label{eq:star_hypertriton_pole}
\end{equation}
STAR's convention assigns negative $f_0$ to the bound doublet channel, opposite to the positive-$a$ convention in Eq.~\eqref{eq:effective_range_amplitude}. The constrained analysis found $f_0({\rm D})=-26.1\pm5.6$ fm and reported $B_\Lambda=0.04^{+0.12}_{-0.03}$ MeV at 95\% confidence level \cite{STAR:2025jwe}. This is a direct precedent for extracting a binding energy from a continuum correlation rather than from an invariant-mass difference. 


\subsection{Refit of the STAR \texorpdfstring{$d\Lambda$}{d-Lambda} Data}

The recently published STAR data permit a proof-of-principle test of how our construction with the assumed source and continuum kernel may describe the data.
We use the 60 public correlation points, comprising 20 $k^*$ bins in each of
the 0--10\%, 10--20\%, and 20--60\% centrality classes, together with the
three published $20\times20$ covariance matrices. Here $k^*$ is the relative
momentum of the $d\Lambda$ constituents and not the momentum of the
hypertriton. As a deliberately minimal universal description, we fit all three
centralities with
\begin{equation}
 C_i(k^*)=B_i+
 \frac{A_i}{\left[(k^*)^2+\gamma^2\right]^2},
 \label{eq:star_contact_refit}
\end{equation}
where the pole scale $\gamma$ is common, while $A_i$ and the constant
normalization $B_i$ are centrality-dependent nuisance parameters.
For comparison with a shallow two-body pole, we define the diagnostic scale
\begin{equation}
 B_{\rm app}=\frac{(\gamma)^2}{2\mu_{d\Lambda}}.
 \label{eq:star_apparent_binding}
\end{equation}
This definition does not by itself establish that $B_{\rm app}$ is the vacuum
hypertriton separation energy.

The full-covariance fit gives
\begin{align}
 \chi^2/{\rm dof}&=64.5/53,\nonumber\\
 \gamma&=0.083~{\rm fm}^{-1}=16.3~{\rm MeV}/c,\nonumber\\
 B_{\rm app}&=0.19~{\rm MeV},\nonumber\\
 0.16&<B_{\rm app}<0.23~{\rm MeV}
 \quad(95\%~{\rm profile}).
 \label{eq:star_contact_result}
\end{align}
The fitted baselines are $(1.001,0.969,1.056)$. The corresponding amplitudes
are
\begin{equation}
 (A_{0-10},A_{10-20},A_{20-60})
 =(0.83,1.08,1.23)\times10^{-3},
 \label{eq:star_contact_amplitudes}
\end{equation}
in the momentum units implied by Eq.~\eqref{eq:star_contact_refit}. Their
increase toward peripheral collisions should not yet be interpreted as a
contact-density measurement because a normalized correlation amplitude also
contains source-volume and channel-weight factors.

\begin{figure*}[t]
 \centering
 \includegraphics[width=0.98\textwidth]
 {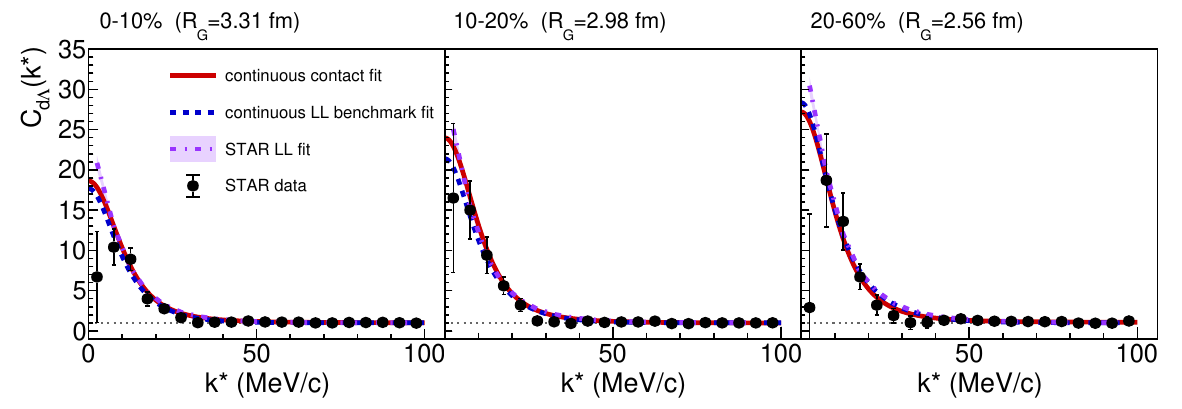}
 \caption{Comparison of the published STAR $d\Lambda$
 correlations in three centrality classes. Black points are the STAR data~\cite{STAR:2025jwe}. 
 Red solid curves are continuous evaluations of the common-$\gamma$ contact
 fit in Eq.~\eqref{eq:star_contact_refit}; blue dashed curves are continuous
 evaluations of an independently fitted analytic LL benchmark. Purple
 dash-dotted curves are STAR's published acceptance-folded LL best fits.
 The benchmark radii are $(3.31,2.98,2.56)$ fm, whereas STAR reports
 $(2.69\pm0.20,2.44\pm0.17,2.08\pm0.19)$ fm. The difference between the two
 LL curves exposes sensitivity to the source radii, scattering parameters,
 effective ranges, and implementation. }
 \label{fig:star_dlambda_contact_ll_comparison}
\end{figure*}

For an analytic benchmark, we also fit the standard neutral nonidentical
Lednick\'y--Lyuboshitz expression with independent Gaussian radii, common
doublet and quartet effective-range parameters, spin weights $1/3$ and $2/3$,
and a real doublet-channel pole. The likelihood is evaluated at the public
bin centers, but the curve in
Fig.~\ref{fig:star_dlambda_contact_ll_comparison} is the resulting continuous
function evaluated on a dense momentum grid. The fit gives
\begin{equation}
 \begin{aligned}
 \chi^2/{\rm dof}&=88/53,\\
 \gamma&=0.050~{\rm fm}^{-1},
 &B&=0.070~{\rm MeV}.
 \end{aligned}
 \label{eq:star_ll_benchmark}
\end{equation}
The radii are $(3.3,3.0,2.6)$ fm and both fitted effective ranges approach
zero. This benchmark is therefore not a reproduction of STAR's published LL
analysis, which fits $R_G$, the doublet and quartet scattering lengths, and
both effective ranges simultaneously within an acceptance-folded Bayesian
calculation. STAR obtains
$R_G=(2.7\pm0.2,2.4\pm0.2,2.1\pm0.2)$ fm and reports
$\chi^2/{\rm dof}=83.4/53$ and
$B_\Lambda=0.04^{+0.12}_{-0.03}$ MeV at 95\% confidence level
\cite{STAR:2025jwe}.

The two LL curves in Fig.~\ref{fig:star_dlambda_contact_ll_comparison} are
retained for comparison: the blue benchmark is evaluated as a continuous post-fit function, whereas the purple curve is STAR's published best-fit result. The comparison in Fig.~\ref{fig:star_dlambda_contact_ll_comparison} therefore
provides both a quantitative and constructive test. A freely normalized contact-only kernel describes the measured enhancement across all three centrality classes, demonstrating that the universal form in Eq.~\eqref{eq:star_contact_refit} can be confronted directly with femtoscopy data. However, this minimal fit absorbs the combined effects of the finite source, doublet and quartet interactions, effective ranges, residual correlations, and detector response into a single effective pole scale. As summarized in Table~\ref{tab:star_contact_binding_results}, the resulting value, $B_{\rm app}\simeq0.19~\mathrm{MeV}$, is approximately five  times the central value reported by STAR~\cite{STAR:2025jwe} and should not be interpreted as an alternative measurement of the vacuum hypertriton separation energy. The most constrained contact description is obtained when the normalized correlation amplitude scales approximately with the inverse source volume,
\begin{equation}
 A_i\propto r_i^{-3}.
 \label{eq:star_inverse_volume_scaling}
\end{equation}
Allowing $A_i\propto r_i^\nu$ gives $\nu=-3.01$, whereas fixing
$\nu=-3$ produces essentially the same minimum. Within this
source-conditioned description, the apparent pole energy changes from
approximately $0.19~\mathrm{MeV}$ in central collisions to
$0.23~\mathrm{MeV}$ in peripheral collisions, apparently in agreement with Wang's construction in Eq.~\eqref{eq:three_way_source_test}. This variation should not be
identified with a centrality dependence of the vacuum hypertriton binding
energy. It characterizes the turnover scale of the effective finite-source
kernel. 

\subsubsection{Hypertriton Bound-Continuum Contact}

The fitted continuum residue can also be expressed as an effective
hypertriton yield by applying the same bound--continuum normalization used
for the universal kernel. For a shallow $d\Lambda$ channel, the normalized
momentum-space kernel is
\begin{equation}
\begin{split}
 \mathcal{K}_{d\Lambda}(k^*;\gamma)
 &=\frac{8\pi\gamma}
 {\left[(k^*)^2+\gamma^2\right]^2},\\
 \int\frac{\dd^3k^*}{(2\pi)^3}
 \mathcal{K}_{d\Lambda}(k^*;\gamma)&=1.
\end{split}
 \label{eq:dlambda_normalized_kernel}
\end{equation}
The coefficient multiplying this kernel is thus the number of
pairs assigned to the corresponding bound channel. Equivalently, the
integrated short-distance pair excess in centrality class $i$ is
\begin{equation}
 N_{H,i}^{\rm corr}
 =
 \int\frac{\dd^3k^*}{(2\pi)^3}\,
 \Delta F_{d\Lambda,i}^{U}(k^*),
 \label{eq:continuum_to_hypertriton_yield}
\end{equation}
where $\Delta F_{d\Lambda,i}^{U}$ denotes the absolute, rather than
unit-normalized, correlated-pair distribution associated with the fitted
universal component. Dividing by the deuteron yield gives
\begin{equation}
 R_{H/d,i}^{\rm corr}
 \equiv
 \frac{N_{H,i}^{\rm corr}}{N_{d,i}}.
 \label{eq:continuum_hypertriton_ratio_definition}
\end{equation}
Thus, the conversion does not identify the fitted pole energy with the
vacuum separation energy. It uses only the integrated residue of the
continuum kernel and the bound--continuum normalization implied by
Eq.~\eqref{eq:dlambda_normalized_kernel}.

Applying this construction to the three STAR centrality classes and
combining the results with their covariance gives the order-of-magnitude
estimate
\begin{equation}
 R_{H/d}^{\rm corr}
 \simeq(1.5\pm0.7)\times10^{-3}.
 \label{eq:star_continuum_hypertriton_ratio}
\end{equation}
The quoted uncertainty propagates the continuum-fit uncertainty within the
minimal contact description. It does not include a complete systematic
uncertainty associated with spin-channel decomposition, residual
correlations, detector-response folding, or conversion of the normalized
correlation into an absolute pair density. Equation~\eqref{eq:star_continuum_hypertriton_ratio}
should therefore be interpreted as a proof-of-principle bound--continuum
consistency estimate, not yet as an independent precision measurement of the
hypertriton production yield.

\subsubsection{Comparison with Bound-State Production}

The continuum-based estimate can be compared with the contact-production
prediction and with independent bound-state measurements. We define
\begin{equation}
 R_{H/d}\equiv
 \frac{\dd N_{{}^{3}_{\Lambda}\mathrm H}/\dd y}
      {\dd N_d/\dd y}.
 \label{eq:hypertriton_deuteron_ratio_definition}
\end{equation}
The five determinations used in this comparison are summarized in
Table~\ref{tab:hypertriton_deuteron_ratios}. When a correction for the
two-body hypertriton decay is required, we use the common convention
\begin{equation}
 \mathrm{B.R.}
 \left(
 {}^{3}_{\Lambda}\mathrm H
 \rightarrow{}^{3}\mathrm{He}+\pi^-
 \right)=0.25.
 \label{eq:hypertriton_branching_convention}
\end{equation}

\begin{table*}[t]
\centering
\caption{
Comparison of hypertriton-to-deuteron ratios inferred from the STAR
$d\Lambda$ continuum correlation, the contact-production calculation of
Braaten, Ingles, and Pickett, and measured bound-state yields. The STAR
top-energy result is an approximate construction from the product
$(H/{}^{3}\mathrm{He})(t/d)$. All branching-corrected
entries use the branching convention in
Eq.~\eqref{eq:hypertriton_branching_convention}.
}
\label{tab:hypertriton_deuteron_ratios}
\small
\setlength{\tabcolsep}{5pt}
\renewcommand{\arraystretch}{1.15}
\begin{tabular}{@{}llllp{0.30\textwidth}@{}}
\toprule
Determination
& Collision system
& $\sqrt{s_{\rm NN}}$
& $R_{H/d}\;(10^{-3})$
& Origin \\
\midrule

STAR continuum
& Au--Au
& $3~\mathrm{GeV}$
& $1.5\pm0.7$
& Integrated residue of the fitted $d\Lambda$ universal kernel \\

Braaten--Ingles--Pickett
& Pb--Pb, 0--10\%
& $2.76~\mathrm{TeV}$
& $1.06\pm0.43$
& Contact-production prediction \cite{Braaten:2024cke} \\

ALICE
& Pb--Pb, 0--10\%
& $2.76~\mathrm{TeV}$
& $1.50\pm0.39$
& Measured hypertriton and deuteron yields
  \cite{ALICE:2015oer,ALICE:2015wav} \\

STAR
& Au--Au, 0--10\%
& $3~\mathrm{GeV}$
& $0.8\pm0.3$
& Hypertriton and deuteron yields measured in the same collision system,
  energy, centrality, and rapidity region
  \cite{STAR:2021orx,STAR:2023uxk} \\

STAR, top energy
& U--U and Au--Au
& $193$--$200~\mathrm{GeV}$
& $0.40\pm0.15$
& Product of the measured $H/{}^{3}\mathrm{He}$ ratio ~\cite{STAR:2023fbc} and 
  top-energy $\mathrm{t}/d$ ratio neglecting isospin difference
  ~\cite{STAR:2022hbp} \\

\bottomrule
\end{tabular}
\end{table*}

For the top-energy STAR estimate, the numerical construction is
\begin{equation}
\begin{split}
 R_{H/d}^{\rm STAR,\,top}
 &=
 \left(
 \frac{N_{{}^{3}_{\Lambda}\mathrm H}}
      {N_{{}^{3}\mathrm{He}}}
 \right)_{\rm STAR}
 \left(
 \frac{N_{{}^{3}\mathrm{He}}}{N_d}
 \right)_{\rm STAR}                                                \\
 &=
 (0.233\pm0.066)
 \left[(1.7\pm0.5)\times10^{-3}\right]                              \\
 &\simeq
 (0.40\pm0.15)\times10^{-3}.
\end{split}
\label{eq:star_top_hypertriton_deuteron_ratio}
\end{equation}
Here the first uncertainty combines the published statistical and
systematic uncertainties in quadrature. This result remains an
order-of-magnitude construction because the two constituent ratios do not
have perfectly matched collision-system and kinematic selections~\cite{STAR:2023fbc,STAR:2022hbp}.

As shown in Table~\ref{tab:hypertriton_deuteron_ratios}, all five
determinations are consistent within their present uncertainties at the
$10^{-3}$ scale. This numerical agreement is nontrivial because the entries
are obtained from different observables. The STAR continuum value is
inferred from the integrated $d\Lambda$ contact residue, the
Braaten--Ingles--Pickett value is a theoretical contact-production
prediction, and the remaining three values are derived from measured
bound-state yields or yield ratios.

The agreement does not establish that the contact-only pole energy
$B_{\rm app}$ is the vacuum hypertriton separation energy. It instead shows
that the integrated strength inferred from the $d\Lambda$ continuum is of
the same order as both the observed hypertriton abundance and the contact
prediction. The comparison is therefore a bound--continuum normalization
test rather than a second determination of the vacuum binding energy.

The STAR comparison at $3~\mathrm{GeV}$ is especially valuable because the
hypertriton and deuteron yields and the $d\Lambda$ correlation were measured
in the same collision system and at the same collision energy. Nevertheless,
the present continuum ratio still relies on converting a normalized
correlation into an absolute pair density and on combining the three
continuum centrality classes. For this reason, it should be regarded as a
proof-of-principle estimate. A definitive test should fit the absolute hypertriton yield and the
$d\Lambda$ correlation simultaneously in matched centrality and acceptance
bins. The measured yield would then fix the contact residue, leaving the
continuum momentum dependence and common pole scale as predictions. Such a
joint analysis could determine whether the present agreement reflects a
common freeze-out contact or results from source, channel, residual-
correlation, and detector effects absorbed by the freely normalized
continuum kernel.

More broadly, the successful description of all three centrality classes by
a common universal kernel demonstrates the organizing power of contact theory
in a strongly interacting and rapidly expanding system. It suggests that
short-distance pair localization, composite production, near-threshold poles,
and continuum correlations can be treated within a common effective
framework, even when the underlying microscopic interactions and freeze-out
dynamics are substantially more complicated. At the same time, the model
dependence of $B_{\rm app}$ cautions against identifying a source-conditioned
effective pole directly with the vacuum binding energy. The STAR example
therefore provides both a realistic validation protocol for the proposed
$K\mu$ analysis and a broader test of bound--continuum contact universality
in expanding strongly interacting matter.

\begin{table*}[t]
\centering
\caption{
Pole scales and apparent binding energies obtained from the different
fits to the STAR $d\Lambda$ correlations. The apparent energies are
calculated from
$B_{{\rm app},i}=(\gamma_i)^2/(2\mu_{d\Lambda})$.
For the contact-kernel fits, these values characterize the turnover of
the fitted correlation kernel and are not automatically the vacuum
hypertriton separation energy.
}
\label{tab:star_contact_binding_results}
\small
\resizebox{\textwidth}{!}{
\begin{tabular}{lcccccc}
\toprule
Fit model
& $\gamma_{0-10}$ 
& $\gamma_{10-20}$
& $\gamma_{20-60}$
& $B_{\rm app}^{0-10}$
& $B_{\rm app}^{10-20}$
& $B_{\rm app}^{20-60}$
\\
& \multicolumn{3}{c}{(fm$^{-1}$)}
& \multicolumn{3}{c}{(MeV)}
\\
\midrule
Common-$\gamma$, independent $A_i$
& $0.083$ & $0.083$ & $0.083$
& $0.19$ & $0.19$ & $0.19$
\\

Radius-constrained, independent $A_i$
& $0.080$ & $0.084$ & $0.090$
& $0.18$ & $0.19$ & $0.23$
\\

Free amplitude power, $\nu=-3.01$
& $0.080$ & $0.084$ & $0.091$
& $0.19$ & $0.21$ & $0.23$
\\

Fixed inverse-volume power, $\nu=-3$
& $0.080$ & $0.084$ & $0.091$
& $0.19$ & $0.21$ & $0.23$
\\

Normalized coalescence power, $\nu=-5$
& $0.079$ & $0.083$ & $0.090$
& $0.17$ & $0.19$ & $0.22$
\\

Common amplitude, $\nu=0$
& $0.081$ & $0.086$ & $0.093$
& $0.18$ & $0.20$ & $0.24$
\\

Extensive Wang-contact power, $\nu=1$
& $0.082$ & $0.086$ & $0.093$
& $0.19$ & $0.20$ & $0.24$
\\

Constant-density power, $\nu=5/2$
& $0.082$ & $0.086$ & $0.094$
& $0.19$ & $0.21$ & $0.24$
\\

Independent $\gamma_i$ and $A_i$
& $0.085$ & $0.070$ & $0.090$
& $0.20$ & $0.16$ & $0.22$
\\
\midrule
Bound-pole LL benchmark
& \multicolumn{3}{c}{$\gamma=0.050$}
& \multicolumn{3}{c}{$B=0.070$ MeV}
\\

STAR published LL analysis
& \multicolumn{3}{c}{Effective-range pole continuation}
& \multicolumn{3}{c}{
$B_\Lambda=0.04^{+0.12}_{-0.03}$ MeV at 95\% CL
}
\\
\bottomrule
\end{tabular}
}
\end{table*}

\section{Domain of validity}

The length $a$ in Eq.~\eqref{eq:a_from_r0} characterizes freeze-out,
not the Coulomb atom. The momentum hierarchy in
Eq.~\eqref{eq:scale_hierarchy} is equivalently $r_0\ll a\ll a_B$ in
coordinate space, where $a_B$ is the physical Bohr radius.

The relevant scales and their physical roles are summarized in
Table~\ref{tab:scales}. Keeping them distinct prevents the effective
freeze-out length $a$ from being mistaken for the physical size of the
Coulomb atom.

\begin{table}[t]
\centering
\caption{Characteristic scales for $K\mu$ production at
$r_0=5~\mathrm{fm}$. Here $k$ is the relative $K\mu$ momentum.}
\label{tab:scales}
\small
\setlength{\tabcolsep}{3.5pt}
\renewcommand{\arraystretch}{1.12}
\begin{tabular}{@{}clp{0.48\columnwidth}@{}}
\toprule
Scale & Value & Physical role \\
\midrule
$r_0$
& $5~\mathrm{fm}$
& Freeze-out localization \\

$a$
& $27.8~\mathrm{fm}$
& Effective contact length \\

$a_B$
& $311~\mathrm{fm}$
& Coulomb Bohr radius \\

$\gamma_C$
& $0.635~\mathrm{MeV}/c$
& Coulomb binding momentum \\

$p_0$
& $7.1~\mathrm{MeV}/c$
& Universal-kernel turnover \\

$\Lambda_{\rm fo}$
& $39.5~\mathrm{MeV}/c$
& Freeze-out resolution \\
\bottomrule
\end{tabular}
\end{table}

The universal kernel applies to the short-distance pair distribution
generated by localization at freeze-out. It does not replace the
long-distance Coulomb Hamiltonian that determines the physical atomic
spectrum, nor does localization alone screen Coulomb propagation from a
measured charged-pair correlation. The observed near-cancellation of the STAR
$K\pi$ attractive--repulsive product provides an empirical control over a
finite momentum interval, as formulated in
Eq.~\eqref{eq:kpi_charge_product}; it is not an assertion that the Coulomb
Hamiltonian has been removed. The measured correlation should therefore
be organized schematically as
\begin{equation}
\begin{split}
 C_{\rm meas}(k)
 ={}&\mathcal{R}_{\rm det}\otimes\mathcal{F}_{\rm src}
 \left[C_{\rm long}^{\rm Coul}(k),
 C_{\rm short}^{U}(k)\right]\\
 &+C_{\rm residual}(k),
\end{split}
 \label{eq:validity_measured_correlation}
\end{equation}
where $\mathcal{F}_{\rm src}$ denotes finite-source averaging and
$\mathcal{R}_{\rm det}$ denotes detector-response folding. The notation in
Eq.~\eqref{eq:validity_measured_correlation} is intentionally not written as
a simple sum: final-state propagation, source averaging, and short-distance
pair production generally act on the pair density or amplitude before the
experimental correlation is formed. An additive separation is justified
only as an approximation after the long-range contribution has been
independently constrained.

The candidate interval $p_0\lesssim k\ll\Lambda_{\rm fo}$ spans
approximately $7.1$--$39.5~\mathrm{MeV}/c$ at $r_0=5~\mathrm{fm}$.
Because it is narrow, the full form in Eq.~\eqref{eq:universal_kernel}
should be tested rather than only its limiting $k^{-4}$ behavior.
Momentum resolution, purity, residual correlations, Coulomb treatment, and
the detailed freeze-out source are therefore quantitatively important.

The contact interpretation also requires the short-distance physics to be
unresolved over the momentum interval of interest. Corrections are expected
when $k\sim\Lambda_{\rm fo}$,
when additional spin or coupled channels become important, or when effective
ranges are no longer small compared with $a$. Such effects generate
higher-order contributions to Eq.~\eqref{eq:universal_kernel} and may
make the effective parameters source- or centrality-dependent. In
particular, a pole scale obtained from a freely normalized correlation fit
is a source-conditioned parameter and need not equal a vacuum binding
momentum.

Observation of the enhanced $K\mu$ atom yield would validate the localized
bound-channel production mechanism and determine its equivalent contact
through Eq.~\eqref{eq:yield_contact_equivalence}.
Using the same contact in the unbound continuum is the additional
universality hypothesis. Its decisive test is not merely whether a smooth
kernel fits the measured line shape, but whether the atom yield fixes both
the continuum residue and the turnover scale without further physics
parameters.

This distinction is illustrated by the STAR $d\Lambda$ analysis. A freely
normalized universal kernel describes the observed low-$k^*$ enhancement,
but its apparent pole scale differs from the independently measured
hypertriton separation energy. The continuum residue nevertheless implies
the order-of-magnitude ratio in
Eq.~\eqref{eq:star_continuum_hypertriton_ratio},
which is consistent with the measured bound-state ratios and the Braaten
contact-production prediction. This agreement is encouraging, but it is
not yet a parameter-free validation because converting a unit-normalized
correlation into an absolute contact requires the absolute uncorrelated-pair
density and its source normalization.

A conclusive test therefore requires a joint yield--correlation analysis in
matched collision systems, centrality classes, and phase-space acceptances.
The bound-state yield must fix the contact residue, the absolute mixed-event
reference must fix the conversion to correlation normalization, and the
charge product must control the dominant Coulomb response. The continuum data
then test the predicted momentum dependence. Agreement of both observables with
a common contact and pole scale would establish bound--continuum
universality. Disagreement would quantify the required higher-order
source-profile, channel, Coulomb-propagation, or detector corrections rather
than invalidating the effective-theory strategy itself.

\section{Conclusion}

We have developed an effective contact theory that connects the production of
Coulomb-bound $K\mu$ atoms to the universal short-distance physics introduced
by Tan. The construction proceeds by replacing Wang's sharp coalescence cutoff
with a smooth universal kernel while preserving the two key predictions of
the original model: the characteristic relative-momentum scale $p_0$ and the
integrated atom yield. The localization scale determines $a$ through
Eq.~\eqref{eq:a_from_r0}, while the atom yield determines $C$ through
Eq.~\eqref{eq:yield_contact_equivalence}.
The replacement therefore introduces neither an independently adjustable
kernel normalization nor an additional matching coefficient.

The $K\mu$ system provides an especially ideal realization of this
construction. For a representative freeze-out radius $r_0=5~\mathrm{fm}$, the
atomic binding momentum, localized production scale, and source resolution
are separated as summarized in Table~\ref{tab:scales}.
This hierarchy separates the long-distance structure of the final Coulomb atom
from the localized process that produces its constituents at freeze-out.
As a result, atom formation can be represented by a short-range effective
kernel even though the asymptotic state is bound by the long-range Coulomb force. A large $K\mu$ atom yield would support the localization-enhanced production predicted by Wang's coalescence model and, within the proposed matching, determine its equivalent freeze-out contact. The atom yield is therefore more than a count of composite particles: it measures the integrated strength of close $K\mu$ pairs created at freeze-out. The clear separation between the atomic, production, and source scales makes this system a particularly direct experimental realization of the Braaten--Tan--Wang approach to weakly bound states. The same contact leads to a testable prediction for $K\mu$ pairs that remain in the continuum: it is fixed by the measured atom yield. The bound and
continuum observables must therefore reproduce the same pole scale and contact normalization. This yield--correlation relation is the central test of the proposed universality.

Our refit of the published STAR $d\Lambda$ correlations
illustrates both the promise and the limitations of a continuum-only analysis.
A common smooth kernel describes the low-$k^*$ enhancement across the measured
centrality classes, and its amplitude scales approximately as the inverse
source volume. Nevertheless, when the contact residue is fitted freely, the
kernel can absorb finite-source effects, doublet and quartet interactions,
effective ranges, residual correlations, and detector response into an
apparent pole. The resulting pole energy can differ substantially
from the independently measured hypertriton separation energy. A more informative comparison follows from translating the continuum normalization into an estimated hypertriton-to-deuteron ratio. It provides the first quantitative indication that composite production and the low-relative-momentum continuum enhancement may be organized by a common freeze-out contact.

The decisive next step is a joint yield--correlation analysis performed in the
same collision system, phase-space acceptance, and centrality classes. The
composite yield should determine the contact residue, while the continuum
correlation tests the predicted momentum dependence and common pole scale.
With the normalization fixed independently, the correlation can no longer
generate an apparent pole merely by absorbing source and detector effects.
Its centrality dependence can then reveal whether the extracted scale is
consistent with a source-independent vacuum state or instead reflects a
source-conditioned structure generated during freeze-out.

The framework thus provides a general strategy for connecting coalescence,
femtoscopy, and near-threshold spectroscopy. It can be applied not only to
$K\mu$ atoms and the hypertriton but also to searches for other loosely bound
states in heavy-ion collisions. A successful yield-normalized continuum test
would demonstrate that bound-state formation and constituent correlations are
complementary manifestations of the same universal short-distance physics,
extending the empirical reach of Tan's contact from equilibrium QED matter
to strongly interacting, rapidly expanding QCD systems.
\begin{acknowledgments}
We thank Drs. Xin Dong and Yu Hu for providing the STAR fit curves, Prof. Kai-jia Sun and members of the STAR Collaboration for valuable discussions. We used Copilot (BNL professional version) for some of the equation format in this draft and ChatGPT-5.6 for producing the first figure draft, table format and for iterating the language of the draft versions. The original ideas of Wang's construction in connection to Tan's universal contact,  to Baym's and Braaten's constructions and the applications to muonic atoms and hypertritons-$d\Lambda$ bound-continuum constraints are solely from the authors of this paper. This work was supported in part by the National Natural Science Foundation of China under Grant Nos. 12361141827 and 12422510, and by the Office of Nuclear Physics within the U.S. Department of Energy Office of Science under Contract DE-FG02-89ER40531 and DE-SC001270. The National Key Research and Development Program of China provided additional support under Contract No. 2022YFA1604900.
W. Zha acknowledges support from the Anhui Provincial Natural Science Foundation (Grant No. 2508085JX002), the Youth Innovation Promotion Association of the Chinese Academy of Sciences, and the Chinese Academy of Sciences under Grant No. YSBR088.

\end{acknowledgments}

\section*{Data Availability}

The published STAR data used in Fig.~\ref{fig:star_dlambda_contact_ll_comparison} are reported in
Ref.~\cite{STAR:2025jwe}. The numerical values of the STAR
fit curves were provided by the authors of that work.
The numerical results and analysis code generated in the
present study are available from the authors upon
reasonable request.

\bibliography{Effective_Contact_Theory_PRX_v31}
\end{document}